\documentclass[11pt]{article}
\usepackage[T1]{fontenc}
\usepackage[utf8]{inputenc}
\usepackage{lmodern}
\usepackage{amsmath,amssymb}
\usepackage{graphicx}
\usepackage{booktabs,longtable,array}
\usepackage{calc}
\usepackage{microtype}
\usepackage{cite}
\usepackage[hidelinks]{hyperref}
\usepackage[margin=1in]{geometry}
\makeatletter
\providecommand{\maxwidth}{\ifdim\Gin@nat@width>\linewidth\linewidth\else\Gin@nat@width\fi}
\makeatother
\title{tSymPerturb converts longitudinal symptom networks into\\time-indexed intervention strategies}
\author{%
Zheng Zhu\textsuperscript{1,4,*}, Junwen Yu\textsuperscript{2,4}, Tiantian Hu\textsuperscript{1,4}, Zhongfang Yang\textsuperscript{3,4}, Jiaqing Wang\textsuperscript{4}\\[0.6em]
\small \textsuperscript{1}School of Nursing, Fudan University, Shanghai, China\\
\small \textsuperscript{2}NYU Shanghai, Shanghai, China\\
\small \textsuperscript{3}School of Nursing, Soochow University, Suzhou, China\\
\small \textsuperscript{4}Yulin AI-Enhanced HealthCare Lab, Shanghai, China\\[0.4em]
\small \textsuperscript{*}Correspondence: \href{mailto:zhengzhu@fudan.edu.cn}{zhengzhu@fudan.edu.cn}}
\date{}

\begin{document}
\maketitle

\begin{abstract}
Longitudinal symptom networks encode directed prediction across
measurement occasions, but outgoing connectivity does not by itself
identify which symptom should be modified, how strongly it should be
changed, or how a perturbation would propagate to later symptoms. We
introduce tSymPerturb, a temporal extension of SymPerturb for
cross-lagged panel networks (CLPNs). The framework separates
source-state operators (temporal virtual knockout and knockdown),
transition operators (directed edge and source-node communication
blocking), and strategy procedures (dosage perturbation, combination
analysis and sequence optimisation). For a two-wave linear CLPN, the
central propagation identity is \(\Delta\mu_2 = B(\mu_1 - \mu_1^{\ast})\), which makes the
source time, outcome time and transition operator explicit. The
formulation also yields three falsification constraints: dose response
is exactly linear under a fixed linear transition model and linear dose
map; independent source-state perturbations are additive at the mean
level; and genuine treatment order is not identified from a single
two-wave transition. In a known 22-node, four-module generating system,
analytical temporal-knockout responses agreed with 250,000-draw Monte
Carlo estimates within 0.0057 standard deviations. Across 200
independently generated datasets, median Spearman correlation with the
population tVPPS ranking increased from 0.76 at n=250 to 0.88 at n=500
and 0.93 at n=1,000; median top-five recovery was 0.60, 0.80 and 0.80,
respectively. Multi-wave simulations showed that target profiles can
change across propagation horizons despite high overall rank
concordance. tSymPerturb therefore converts longitudinal network
structure into auditable, time-indexed intervention hypotheses while
retaining the distinction between prediction and causal treatment
effects.
\end{abstract}

\section{1 Introduction}\label{introduction}

Network models have changed how multivariate symptom systems are
represented. Rather than treating symptoms only as interchangeable
indicators of a latent disorder, network models represent symptoms as
mutually related components whose structure can be studied at a single
occasion or across time \cite{ref1,ref2,ref3,ref4}. Cross-lagged panel networks extend
this logic to panel data by estimating how a symptom at an earlier
occasion predicts itself and other symptoms at a later occasion,
conditional on the remaining source variables \cite{ref3,ref4}. This
directional representation is appealing for symptom management because
it appears to provide a temporal basis for identifying candidate
intervention targets.

However, directional prediction and intervention value are not the same
quantity. Outgoing strength or expected influence describes the fitted
transition structure; it does not quantify how much later symptom burden
would change if a baseline symptom were modified, whether a feasible
partial change would preserve the same ranking, whether a target acts
mainly through persistence or cross-symptom spillover, or which directed
pathways account for the predicted benefit. The distinction becomes
especially important when autoregressive paths are large, when the
source symptoms have different modifiable ranges, or when positive and
negative cross-lagged paths coexist.

The original SymPerturb framework formalised virtual perturbation for
cross-sectional symptom networks by separating primitive perturbation
operators from procedures that evaluate intensity, combinations and
target sequences. Its central limitation is also explicit: an undirected
cross-sectional graph does not contain temporal ordering, and a
statistical perturbation of that graph is not an identified causal
intervention. Longitudinal networks provide a different mathematical
object. A fitted CLPN is a directed mapping from a source state at T1 to
an outcome state at T2. The virtual-perturbation question can therefore
be written as a forward propagation query: what does the fitted
transition model predict at T2 after a prespecified modification of the
T1 state or of the T1\(\rightarrow\)T2 transition process?

This temporal formulation does not automatically solve causal
identification. Cross-lagged coefficients remain regression parameters
that can be affected by stable between-person differences, measurement
error, omitted causes and the selected time interval \cite{ref5,ref6}. Wysocki
and colleagues explicitly recommend interpreting CLPN paths in
predictive rather than causal terms \cite{ref3}. tSymPerturb therefore uses
intervention language only to define transparent model operations. The
output is a model-implied target hypothesis that requires longitudinal
triangulation, measured target engagement and experimental validation
before it can be interpreted as a treatment effect.

Here we introduce tSymPerturb as a methods framework for longitudinal
symptom networks. We first define temporal state and transition
operators, then derive their response functions in a two-wave linear
CLPN, distinguish total downstream benefit from cross-symptom spillover,
formalise seven non-redundant target-level utility outcomes, and extend
the framework to combinations and multi-wave sequences. We then use a
known 22-node, four-module transition system to verify the analytical
equations, quantify finite-sample recovery and expose three conditions
under which seemingly attractive findings---nonlinear dosage curves,
additive synergy and two-wave treatment-order claims---are
mathematically unsupported by the reference model.

\section{2 The tSymPerturb
architecture}\label{the-tsymperturb-architecture}

tSymPerturb is organised into a source-state layer, a transition layer
and a strategy layer (Fig. 1). The source-state layer contains temporal
virtual knockout (t-vKO), which anchors a T1 symptom to a prespecified
state, and temporal virtual knockdown (t-vKD), which produces a partial
movement toward that anchor. Temporal virtual dosage perturbation
(t-vDP) evaluates a response curve generated by a state operator. The
transition layer contains directed edge-level communication blocking and
source-node-centred communication blocking, which attenuate one T1\(\rightarrow\)T2
coefficient or all outgoing coefficients from a selected source node.
The strategy layer contains combination perturbation and sequence
optimisation. The latter has two distinct meanings: target-acquisition
order can be studied in a two-wave decision problem, whereas repeated
temporal intervention order requires at least a multi-step transition
model or an explicit stationarity assumption.

\begin{figure}[htbp]
\centering
\includegraphics[width=\linewidth]{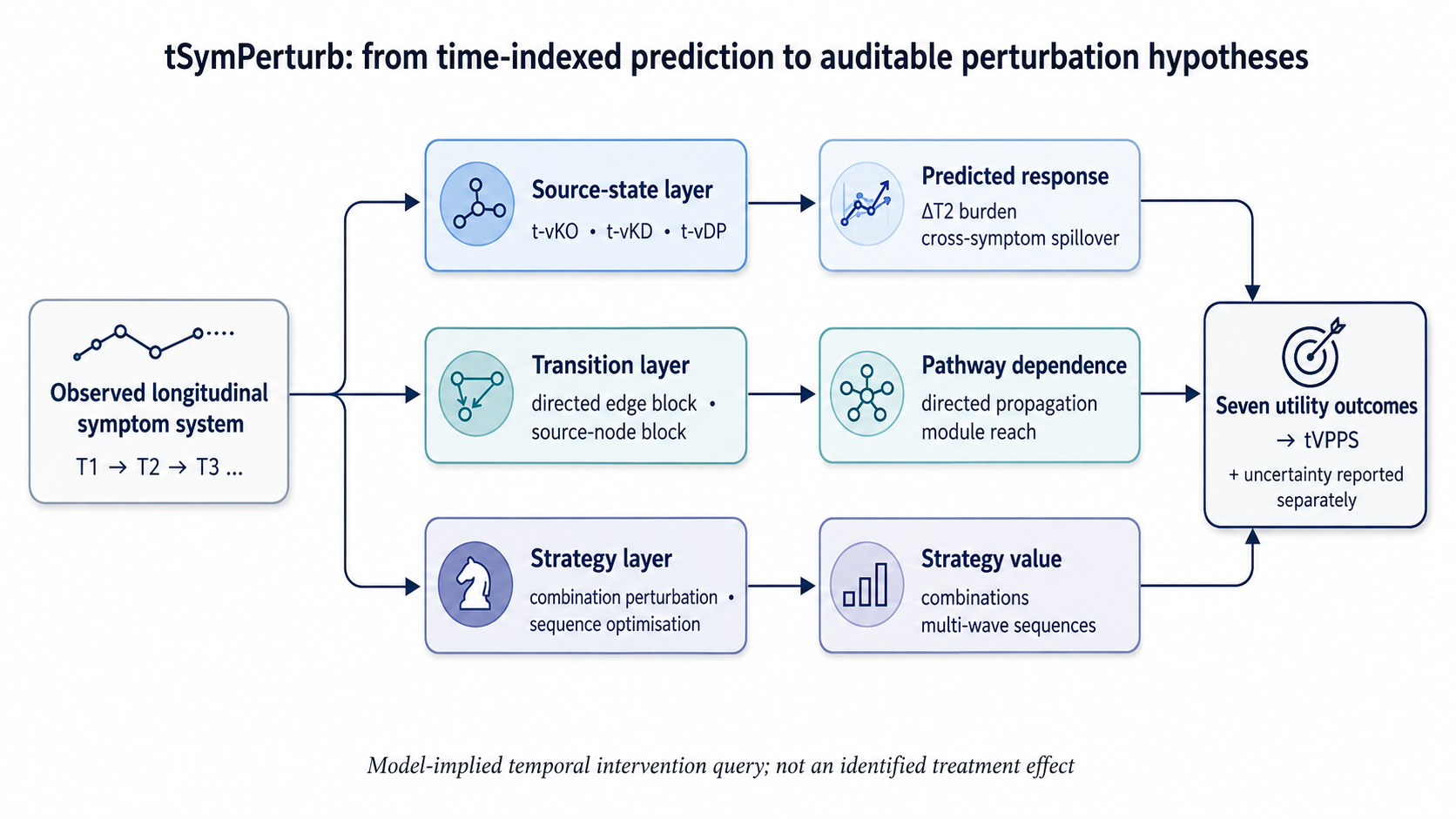}
\caption{The tSymPerturb workflow. Source-state operators alter the symptom state at the source occasion; transition operators alter the fitted T1\(\rightarrow\)T2 mapping; strategy procedures combine or order candidate interventions. Seven target-level utility outcomes are used for within-analysis prioritisation, while uncertainty is reported separately. All quantities are model-implied and do not represent identified treatment effects.}
\label{fig:workflow}
\end{figure}

\noindent\textbf{Table 1.} Components of tSymPerturb and the questions they answer.

\begin{longtable}[]{@{}
  >{\raggedright\arraybackslash}p{(\columnwidth - 4\tabcolsep) * \real{0.2139}}
  >{\raggedright\arraybackslash}p{(\columnwidth - 4\tabcolsep) * \real{0.2830}}
  >{\raggedright\arraybackslash}p{(\columnwidth - 4\tabcolsep) * \real{0.5031}}@{}}
\toprule\noalign{}
\endhead
\bottomrule\noalign{}
\endlastfoot
\textbf{Component} & \textbf{Reference action} & \textbf{Mathematical
role and primary question} \\
Temporal virtual knockout (t-vKO) & Anchor source symptom i at \(c_i\) &
Primitive state operator: what is the predicted downstream response to
complete source-state anchoring? \\
Temporal virtual knockdown (t-vKD) & Partially move source symptom i
toward \(c_i\) & Primitive state operator: what response follows a feasible
partial source change? \\
Temporal virtual dosage perturbation (t-vDP) & Evaluate \(G_i(d)\) over
\(d\in[0,1]\) & Intensity-response procedure: is any apparent nonlinearity
attributable to the specified model rather than the CLPN alone? \\
Directed edge communication block & \(B_{ji}^{\ast}=(1-q)B_{ji}\) & Primitive transition
operator: how dependent is a selected prediction on one directed
pathway? \\
Source-node communication block & Attenuate the outgoing column \(B_{\cdot i}\) &
Primitive transition operator: how much finite-horizon propagation
depends on one source node? \\
Combination perturbation & Apply a joint source-state operator to S &
Joint-target construction: what incremental value does an additional
target contribute beyond the better single target? \\
Sequence optimisation & Optimise ordered target acquisition or repeated
interventions & Decision procedure: which feasible order maximises a
prespecified temporal objective under explicit assumptions? \\
\end{longtable}

\section{3 A temporal source-state intervention with virtual knockout as
an
endpoint}\label{a-temporal-source-state-intervention-with-virtual-knockout-as-an-endpoint}

Let \(X_1\in\mathbb{R}^p\) and \(X_2\in\mathbb{R}^p\) denote the symptom vectors at two consecutive
occasions. The reference CLPN is

\[X_{2} = a + BX_{1} + \varepsilon,\quad\quad E(\varepsilon) = 0,\quad\quad Var(\varepsilon) = \Psi.\quad\quad(1)\]

where \(B_{ji}\) is the coefficient from source symptom i at T1 to outcome
symptom j at T2. Diagonal elements represent autoregressive persistence
and off-diagonal elements represent cross-symptom prediction. The
baseline mean is \(\mu_2=a+B\mu_1\). A state perturbation changes the source
distribution while keeping B fixed. For a target set S with clinically
meaningful anchors \(c_S\), a general location-scale map is

\[X_{1,S}^{(d)} = c_{S} + D_{\mu}(d)\left( \mu_{1,S} - c_{S} \right) + D_{\sigma}(d)\left( X_{1,S} - \mu_{1,S} \right).\quad\quad(2)\]

with non-target source variables unchanged. The post-perturbation
follow-up mean is \(\mu_2(d)=a+B\mu_1(d)\), so the model-implied downstream
improvement is

\[R_{S}(d) = \mu_{2} - \mu_{2}^{(d)} = B\left\lbrack \mu_{1} - \mu_{1}^{(d)} \right\rbrack.\quad\quad(3)\]

Equation (3) is the central temporal propagation identity. It replaces
the cross-sectional re-equilibration query with an explicitly
time-indexed forward mapping. For one target i anchored at \(c_i\), the t-vKO
mean response is

\[R_{i}^{vKO} = B_{\cdot i}\left( \mu_{1,i} - c_{i} \right).\quad\quad(4)\]

This response depends jointly on the modifiable distance from the source
state to the anchor and on the outgoing transition profile. Unlike exact
knockout in a cross-sectional covariance model, t-vKO does not require
re-estimating a covariance matrix containing a zero-variance column. The
fitted transition model is estimated once and then queried. Re-fitting a
CLPN after replacing a predictor by a constant would answer a different
question and may create rank deficiency.

\section{4 Seven components and longitudinal-specific
constraints}\label{seven-components-and-longitudinal-specific-constraints}

\subsection{Temporal knockdown and dosage
perturbation}\label{temporal-knockdown-and-dosage-perturbation}

A partial knockdown uses \(d\in(0,1)\) to move the source state toward its
anchor. Under the linked reference map \(X_{1,i}^{\ast}=c_i+(1-d)(X_{1,i}-c_i)\), the mean
source improvement is \(d(\mu_{1,i}-c_i)\), giving

\[R_{i}(d) = dB_{\cdot i}\left( \mu_{1,i} - c_{i} \right).\quad\quad(5)\]

Consequently, when B is fixed and the outcome functional is linear,
\(G_i(d)=dG_i(1)\). Dose efficiency and low-dose responsiveness are then
algebraically equivalent to efficacy. t-vDP is therefore retained as an
intensity-response procedure but these redundant summaries are not
counted as separate dimensions of the confirmatory tVPPS. Thresholds or
saturation require additional structure, such as bounded measurement, a
nonlinear dose map, interactions, state-dependent coefficients or
nonlinear transition functions.

\subsection{Directed communication
blocking}\label{directed-communication-blocking}

State perturbation changes X\textsubscript{1}; communication blocking changes B. For a
directed pathway \(i(T1)\rightarrow j(T2)\), a block fraction \(q\in[0,1]\) gives

\[B^{\ast (i \rightarrow j)} = B - qB_{ji}e_{j}e_{i}^{\top}.\quad\quad(6)\]

A source-node block attenuates the outgoing transition column \(B_{\cdot i}\). In
individual-level or clinically anchored predictions, this operator
quantifies dependence of future symptom burden on the selected
predictive pathway. When predictors are centred, communication-block
effects must be evaluated at an explicit reference state; otherwise the
average signed change can be zero by construction. For target-level
scoring, the reference implementation uses the relative loss of a
finite-horizon propagation functional after 80\% source-node blocking.

\subsection{Combination perturbation}\label{combination-perturbation}

For two independently perturbed source symptoms i and k under the linear
reference model, the joint mean response is \(R_{\{i,k\}}=R_i+R_k\). Thus the
additive interaction contrast \(G_{\{i,k\}}-G_i-G_k\) is exactly zero. A two-wave
linear CLPN cannot generate statistical synergy merely because two
baseline states are modified together. tSymPerturb instead reports
combination value relative to the better single target and retains the
additive contrast as a diagnostic. Non-zero non-additivity must arise
from an explicitly nonlinear element of the model, outcome function or
state-update rule.

\subsection{Sequence optimisation}\label{sequence-optimisation}

Two-wave and multi-wave sequence questions are deliberately separated. A
two-wave model can rank the order in which candidate targets are added
to a constrained intervention set, but it contains only one observed
transition and therefore does not identify a biological treatment
sequence. Repeated temporal sequencing requires additional observed
waves or a stated stationarity assumption under which the fitted
transition operator is reused. This distinction prevents an optimisation
algorithm from being misreported as evidence of treatment timing.

\section{5 Seven utility outcomes, uncertainty and
tVPPS}\label{seven-utility-outcomes-uncertainty-and-tvpps}

For target i, let \(\Delta_{j\leftarrow i}(d)\) denote the standardised predicted improvement
in outcome j after a source-state perturbation. tSymPerturb
distinguishes total downstream efficacy, which includes the
autoregressive outcome, from cross-symptom spillover, which excludes it.
Seven non-redundant utility outcomes are retained for the reference
two-wave tVPPS (Table 2). Robustness is treated as uncertainty rather
than utility and is reported separately.

\noindent\textbf{Table 2.} Seven target-level utility outcomes used in the reference tVPPS.

\begin{longtable}[]{@{}
  >{\raggedright\arraybackslash}p{(\columnwidth - 4\tabcolsep) * \real{0.2012}}
  >{\raggedright\arraybackslash}p{(\columnwidth - 4\tabcolsep) * \real{0.3145}}
  >{\raggedright\arraybackslash}p{(\columnwidth - 4\tabcolsep) * \real{0.4842}}@{}}
\toprule\noalign{}
\endhead
\bottomrule\noalign{}
\endlastfoot
\textbf{Outcome} & \textbf{Reference definition} &
\textbf{Interpretation} \\
Downstream efficacy & Mean \(\Delta_{j\leftarrow i}(1)\) over all T2 outcomes & Overall
predicted follow-up burden reduction, including persistence of the
target itself. \\
Cross-symptom spillover & Mean \(\Delta_{j\leftarrow i}(1)\), \(j\neq i\) & Predicted benefit that
propagates beyond the target symptom. \\
Breadth & Share of \(j\neq i\) with \(\Delta_{j\leftarrow i}\geq 0.05\) SD & Proportion of subsequent
symptoms exceeding a prespecified improvement threshold. \\
Cross-module reach & Share of other modules with mean improvement \(\geq 0.03\)
SD & Extent to which the predicted response reaches other symptom
domains. \\
Communication-block value & Relative loss of a finite-horizon
\textbar B\textbar-propagation functional after 80\% source blocking &
Dependence of the directed transition system on the source node. \\
Combination value & Mean incremental pair value over a prespecified
partner set & Additional value beyond the better single target; not
causal synergy. \\
Spillover fraction & Positive non-target response divided by total
positive response & Fraction of beneficial response that is distributed
beyond autoregressive target persistence. \\
\end{longtable}

Each raw outcome \(R_{im}\) is direction-aligned and min--max normalised across
the prespecified candidate set. If a dimension is constant it
contributes a neutral score of 50. The reference temporal virtual
perturbation priority score is

\[{tVPPS}_{i} = \frac{\sum_{m}^{}\omega_{m}{score}_{im}}{\sum_{m}^{}\omega_{m}},\quad\quad\omega_{m} \geq 0.\quad\quad(7)\]

Equal weights are used only for methodological verification. They are
not patient utilities and do not make tVPPS transportable across
cohorts. Applied work should report the seven-dimensional profile beside
tVPPS and pre-specify any clinically informed weights. Because
normalisation is within the analysed candidate set, tVPPS is a relative
prioritisation score rather than an absolute treatment-benefit scale.

\section{6 Internal computational verification and finite-sample
recovery}\label{internal-computational-verification-and-finite-sample-recovery}

We generated a known 22-node transition system with four symptom
modules. The matrix contained 53 non-zero off-diagonal transitions, of
which 47 were positive and 6 negative, together with heterogeneous
autoregressive coefficients ranging from 0.43 to 0.71. The spectral
radius of B was 0.827, allowing a stable stationary extension for the
multi-wave stress test. Baseline means ranged from 1.44 to 2.64,
baseline standard deviations from 0.63 to 0.83, and residual standard
deviations from 0.38 to 0.52. Population outcomes were calculated from
the known generating parameters before any data were sampled.

Analytical t-vKO responses closely matched direct simulation. Across 22
targets and 250,000 Monte Carlo draws per target, the maximum absolute
discrepancy in any standardised T2 outcome was 0.0057 SD. Target-level
analytical and Monte Carlo efficacy values had Pearson r=0.99991, with a
maximum absolute efficacy discrepancy of 0.0013 SD (Fig. 2a). This
verifies the forward propagation implementation under model
compatibility; it does not establish the clinical validity of the target
anchors or the transition model.

Finite-sample recovery improved monotonically with sample size. Across
200 independently generated datasets at each n, the median Spearman
correlation between estimated and population tVPPS ranks was 0.76 at
n=250, 0.88 at n=500 and 0.93 at n=1,000 (Fig. 2b and Table 3). The
corresponding 95\% Monte Carlo intervals were 0.58--0.88, 0.74--0.95 and
0.85--0.98. Median top-five recovery increased from 0.60 to 0.80 and
0.80, while tVPPS RMSE decreased from 13.94 to 9.47 and 6.39 points on
the 0--100 scale (Fig. 2c).

Outcome-specific recovery was heterogeneous (Fig. 2d). Efficacy,
cross-symptom spillover, communication-block value, combination value
and spillover fraction were recovered well by n=500, with median rank
correlations of 0.90 or higher for all except cross-module reach.
Threshold-based breadth was the least stable component, with median rank
recovery of 0.41, 0.60 and 0.80 across n=250, 500 and 1,000. This
pattern supports reporting the raw utility profile and complete-pipeline
uncertainty rather than treating the composite rank as error-free.

\begin{figure}[htbp]
\centering
\includegraphics[width=\linewidth]{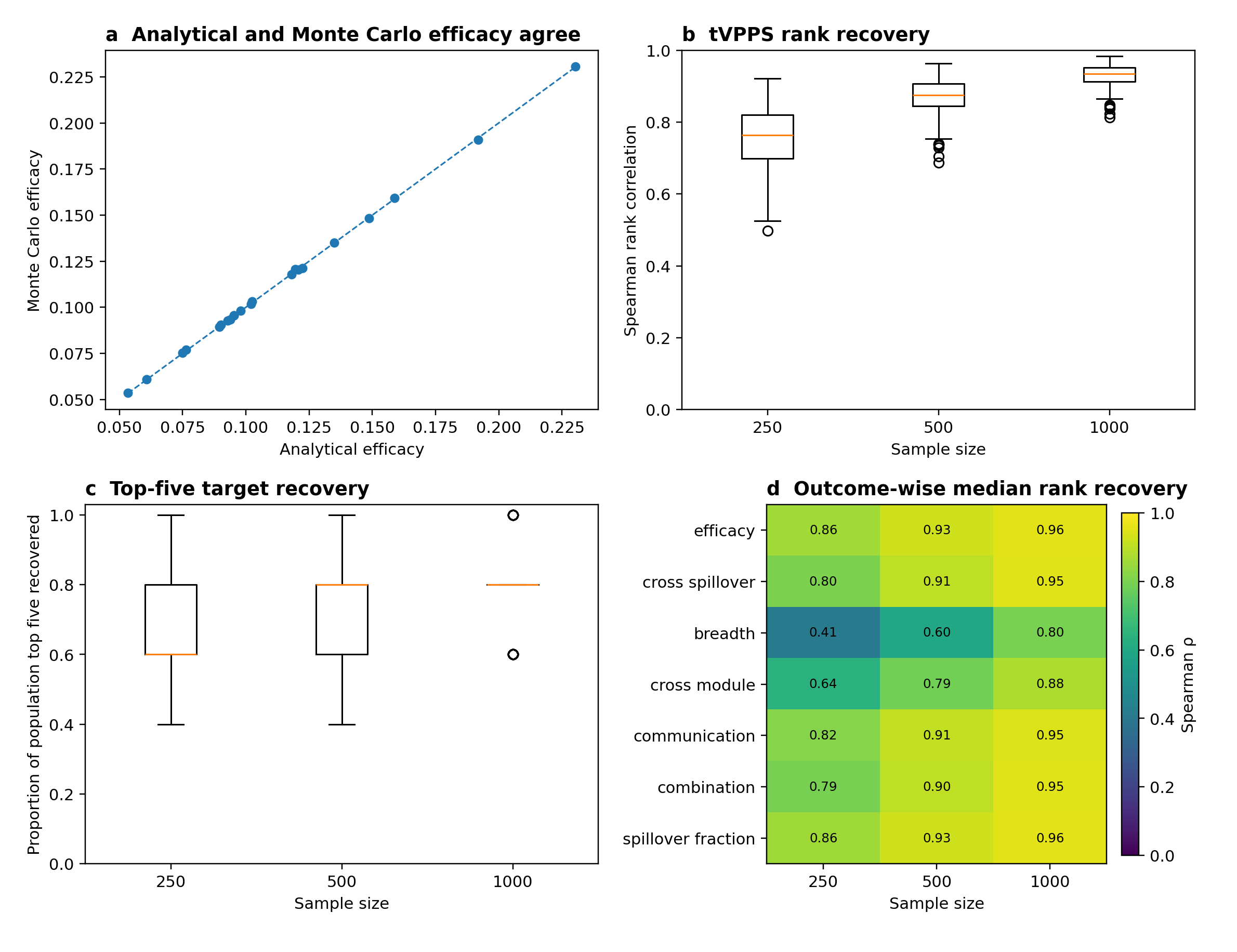}
\caption{Internal verification and finite-sample recovery in model-compatible longitudinal networks. a, Analytical versus Monte Carlo t-vKO efficacy for 22 targets. b, Distribution of tVPPS rank correlation across 200 independently generated datasets at each sample size. c, Proportion of the population top five recovered. d, Median outcome-specific rank recovery. Monte Carlo intervals quantify simulation-to-simulation variability under the specified generating system and are not patient-level confidence intervals.}
\label{fig:verification}
\end{figure}

\noindent\textbf{Table 3.} Finite-sample recovery of the seven-dimension tVPPS. Intervals are the 2.5th and 97.5th percentiles across 200 independently generated datasets.

\begin{longtable}[]{@{}
  >{\raggedright\arraybackslash}p{(\columnwidth - 6\tabcolsep) * \real{0.1446}}
  >{\raggedright\arraybackslash}p{(\columnwidth - 6\tabcolsep) * \real{0.2956}}
  >{\raggedright\arraybackslash}p{(\columnwidth - 6\tabcolsep) * \real{0.2956}}
  >{\raggedright\arraybackslash}p{(\columnwidth - 6\tabcolsep) * \real{0.2641}}@{}}
\toprule\noalign{}
\endhead
\bottomrule\noalign{}
\endlastfoot
\textbf{Sample size} & \textbf{Median tVPPS \(\rho\) (95\% MC interval)} &
\textbf{Median top-five recovery (95\% MC interval)} & \textbf{Median
RMSE, 0--100 (95\% MC interval)} \\
250 & 0.76 (0.58--0.88) & 0.60 (0.40--0.80) & 13.94 (9.81--18.46) \\
500 & 0.88 (0.74--0.95) & 0.80 (0.60--1.00) & 9.47 (6.58--13.11) \\
1,000 & 0.93 (0.85--0.98) & 0.80 (0.60--1.00) & 6.39 (4.10--9.25) \\
\end{longtable}

At the population level, Fatigue ranked first (tVPPS=100.0), followed by
Rumination (84.6), Anxiety (61.3), Pain (61.3) and Anhedonia (52.2). The
ranking was not reducible to baseline symptom burden: Pain had high
total efficacy but a smaller spillover fraction than Fatigue or
Rumination, whereas Rumination combined broad cross-module reach with
strong transition dependence. These labels identify the intentionally
constructed synthetic nodes and must not be interpreted as clinical
recommendations.

\section{7 Temporal validity, multi-wave propagation and benchmark
contrasts}\label{temporal-validity-multi-wave-propagation-and-benchmark-contrasts}

The reference linear model satisfied the two main algebraic
falsification checks to numerical precision. Across all 22 targets and
six dose levels, the maximum deviation from \(G_i(d)=dG_i(1)\) was 2.78e-17.
Across all 231 target pairs, the maximum absolute deviation from joint
additivity was 2.78e-17. Thus, any visibly nonlinear dosage curve or
non-additive pair effect produced by the unbounded two-wave reference
implementation would indicate a coding error or the introduction of an
additional nonlinear operation rather than a property learned from the
CLPN itself (Fig. 3a,b).

The multi-wave extension produced a distinct longitudinal result.
Reusing the known stable transition matrix over four waves changed the
relative target profiles even though ranks remained broadly concordant.
The one-step and four-step efficacy ranks correlated at \(\rho\)=0.88; the
correlation between one-step efficacy and discounted four-wave
cumulative efficacy was \(\rho\)=0.94. Fatigue was the highest one-step target,
whereas Rumination became the highest four-step target; Insomnia entered
the top five after the first transition (Fig. 3c). This illustrates why
one-step outgoing influence and persistent longitudinal leverage are not
interchangeable.

Under an exploratory stationary repeated-intervention simulation
restricted to the eight highest population tVPPS candidates, the highest
discounted two-step utility was obtained for Fatigue\(\rightarrow\)Pain (U=0.592),
compared with U=0.527 for the reverse order. The order sensitivity for
this pair was 0.065. Because the same transition matrix was reused
beyond the observed two-wave interval, this result is a strategy
simulation under stationarity, not evidence that the sequence should be
used clinically. In the pure two-wave target-acquisition problem without
state updating or constraints, cumulative utility was order invariant as
required.

tVPPS was associated with conventional longitudinal centrality but was
not identical to it. Population tVPPS rank correlated with absolute
outgoing-strength rank at \(\rho\)=0.80 and signed outgoing expected-influence
rank at \(\rho\)=0.89; only 60\% of the top five targets overlapped with either
outgoing-strength top five. Baseline burden alone had a much weaker
association with tVPPS (\(\rho\)=0.16; top-five overlap 40\%). The contrast
shows that tSymPerturb combines source modifiability, directed response,
module reach and strategy value rather than reproducing a single
centrality statistic (Fig. 3d).

\begin{figure}[htbp]
\centering
\includegraphics[width=\linewidth]{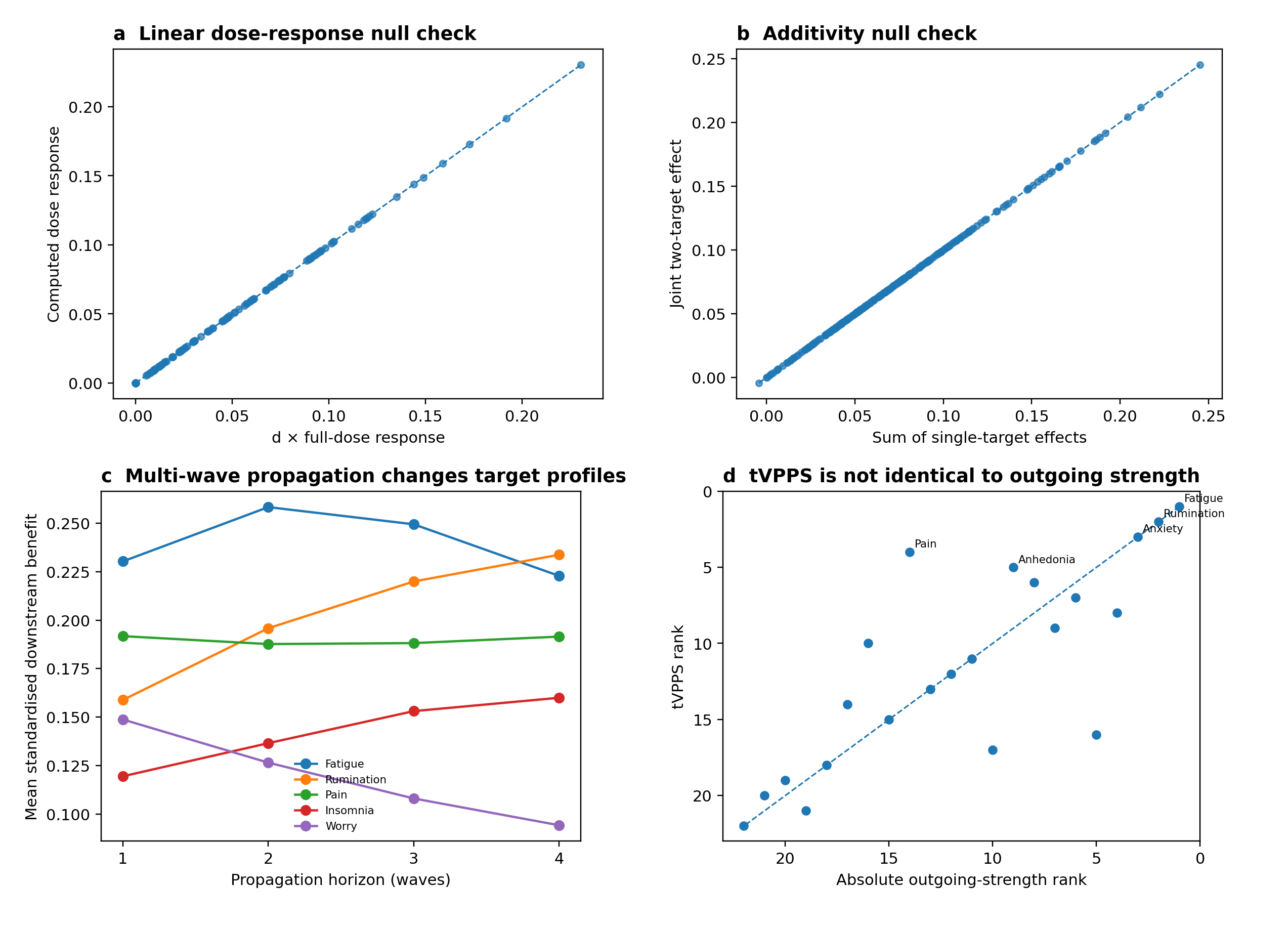}
\caption{Longitudinal-specific validity checks and contrasts. a, The unbounded linear reference model lies on the analytical dose-response identity. b, Joint two-target state perturbations lie on the additive identity. c, Model-implied target efficacy across four repeated transition horizons under the stationary extension. d, Population tVPPS rank versus absolute outgoing-strength rank. The multi-wave and sequence analyses are simulations under the specified transition model, not identified treatment policies.}
\label{fig:validity}
\end{figure}

\section{8 Discussion}\label{discussion}

tSymPerturb extends virtual perturbation from contemporaneous symptom
structure to explicitly time-indexed prediction. Its central
contribution is not another centrality statistic. The framework
specifies the object being perturbed, the temporal location of that
perturbation, the rule by which the fitted system is updated and the
outcome functional used to judge the response. Source-state operators
ask what the fitted CLPN predicts after a change in the earlier symptom
state. Transition operators ask how much later prediction depends on a
directed pathway. Combination and sequence procedures then construct
intervention strategies from those primitive operations.

The mathematical formulation clarifies several quantities that are easy
to overinterpret in applied longitudinal network studies. First, a T1
symptom with large outgoing coefficients is not automatically the best
intervention target because the model-implied response also depends on
its distance from a clinically meaningful anchor and on the outcomes
reached by those coefficients. Second, autoregressive persistence and
cross-symptom propagation answer different questions. A target can
reduce its own future burden strongly while having little spillover, or
it can have a smaller total effect but a larger fraction of benefit
distributed across other symptoms. Third, communication blocking is a
transition-model operation. It should not be interpreted as evidence
that symptoms literally transmit a biological signal.

The simulation results support the internal coherence of the framework
while also defining sample-size limitations. Analytical and Monte Carlo
responses agreed closely, and population rankings were increasingly
recovered as n increased. Nevertheless, at n=250 the 95\% Monte Carlo
interval for tVPPS rank recovery extended from 0.58 to 0.88 and top-five
recovery could be as low as 0.40. Breadth was particularly unstable
because a threshold converts small coefficient errors into discrete
changes in the count of affected symptoms. These findings argue against
reporting only a single target rank. Applied analyses should propagate
uncertainty through estimation, perturbation, normalisation and ranking,
then report rank distributions and top-k selection probabilities beside
the point estimate.

Three null results are as important as the positive validation. In the
reference linear CLPN, dosage response is linear, independent source
perturbations are additive and a two-wave model contains no empirical
information about repeated treatment order. These are not inconveniences
to be hidden by a more elaborate score. They are falsification
constraints. A sigmoid dosage curve, pairwise synergy or strong order
sensitivity can be scientifically meaningful only when the analysis
explicitly introduces the mechanism that can generate it---for example
bounded outcomes, nonlinear transitions, interactions, state-dependent
coefficients, intervention constraints or observed multi-wave
transitions.

The multi-wave extension illustrates the additional information that
becomes available when a transition process is iterated. In the
synthetic system, one-step and four-step ranks were similar but not
identical; a target that was not among the strongest immediate
candidates became more prominent at later horizons. This distinction is
important for chronic symptom management, where an intervention may have
modest immediate benefit yet alter a downstream cascade over several
assessment intervals. However, repeating one estimated B matrix assumes
temporal stationarity. When three or more empirical waves are available,
wave-specific matrices \(B_t\) should be used unless invariance is supported.

Several limitations define the scope of the present claims. The
verification uses a continuous Gaussian generating system with a sparse
linear transition matrix and a fixed lasso penalty. It does not
establish performance for strongly ordinal or zero-inflated symptoms,
non-random dropout, measurement non-invariance, latent time-varying
confounding, heterogeneous person-specific dynamics or high-dimensional
p/n regimes. The seven tVPPS dimensions are methodological utilities
rather than validated clinical preference weights. The target anchor,
burden weights and thresholds for breadth and module reach are analysis
choices. Most importantly, temporal precedence within a CLPN does not
identify causal treatment effects \cite{ref5,ref6,ref16}.

A credible translational pathway should therefore proceed in stages. The
first is expanded computational validation across data types,
estimators, missingness mechanisms, nonlinearity and time-varying
transition matrices. The second is longitudinal triangulation, asking
whether naturally occurring or treatment-induced changes in high-ranked
source symptoms precede the downstream pattern predicted by tSymPerturb.
The third is experimental validation with measured target engagement,
followed by prospective decision studies that incorporate safety,
feasibility, cost, equity and patient preference. Used within these
boundaries, tSymPerturb can narrow a target set and make temporal
intervention assumptions testable; it cannot replace a trial.

\section{9 Methods}\label{methods}

\subsection{9.1 Reference two-wave transition
model}\label{reference-two-wave-transition-model}

Let \(X_{1} = \left( X_{1,1},\ldots,X_{1,p} \right)^{\top}\) and
\(X_{2} = \left( X_{2,1},\ldots,X_{2,p} \right)^{\top}\) denote the same
\(p\) symptoms measured at two ordered occasions. The reference
cross-lagged panel network (CLPN) is written as a multivariate
transition model

\[X_{2} = a + BX_{1} + \varepsilon,\quad\quad E(\varepsilon) = 0,\quad\quad Var(\varepsilon) = \Psi,\quad\quad Cov\left( X_{1},\varepsilon \right) = 0.\quad\quad(8)\]

The coefficient \(B_{ji}\) is the directed path from source symptom
\(i\) at \(T1\) to outcome symptom \(j\) at \(T2\). The diagonal
\(B_{ii}\) contains autoregressive persistence and the off-diagonal
elements contain cross-symptom prediction. If
\(E\left( X_{1} \right) = \mu_{1}\) and
\(Var\left( X_{1} \right) = \Sigma_{1}\), the first two moments at
follow-up follow directly from linear propagation:

\[\mu_{2} = E\left( X_{2} \right) = a + B\mu_{1},\quad\quad(9)\]

\[\Sigma_{2} = Var\left( X_{2} \right) = B\Sigma_{1}B^{\top} + \Psi.\quad\quad(10)\]

The source--outcome cross-covariance is

\[Cov\left( X_{1},X_{2} \right) = \Sigma_{1}B^{\top}.\quad\quad(11)\]

Equations (9)--(11) are important because tSymPerturb operates on an
already fitted transition operator. A state perturbation changes the
distribution entering \(B\); a communication-block perturbation changes
\(B\) itself. The two operations therefore answer different estimands
and should not be implemented by repeatedly refitting the CLPN after
every hypothetical intervention.

\subsection{9.2 Clinical anchors and scale
transformation}\label{clinical-anchors-and-scale-transformation}

Every symptom was oriented so that larger values represented worse
states. Let \(c_{i}\) denote a clinically interpretable source-state
anchor for symptom \(i\). The anchor is defined on the original
measurement scale before standardisation. If a fitted CLPN uses

\[Z_{1,i} = \frac{X_{1,i} - \mu_{1,i}}{\sigma_{1,i}},\]

then the corresponding anchor on the fitted scale is

\[c_{i,z} = \frac{c_{i} - \mu_{1,i}}{\sigma_{1,i}}.\quad\quad(12)\]

Thus, when symptom absence is coded as \(c_{i} = 0\), a standardised
value of zero is generally \emph{not} a knockout state; it is the sample
mean. This transformation was repeated inside each finite-sample
replicate so that sampling variability in the estimated mean and
standard deviation propagated into the perturbation analysis.

\subsection{9.3 General temporal location--scale state
intervention}\label{general-temporal-locationscale-state-intervention}

Let \(S\) be a target set and let \(K\) denote the complementary set of
non-target source symptoms. A general temporal intervention may alter
the target mean and target scale separately. For target vector
\(X_{1,S}\), define

\[X_{1,S}^{(d)} = c_{S} + D_{\mu}(d)\left( \mu_{1,S} - c_{S} \right) + D_{\sigma}(d)\left( X_{1,S} - \mu_{1,S} \right),\quad\quad(13)\]

where \(D_{\mu}(d)\) and \(D_{\sigma}(d)\) are diagonal intervention
maps. Baseline recovery requires \(D_{\mu}(0) = D_{\sigma}(0) = I\). An
exact anchor intervention requires \(D_{\mu}(1) = D_{\sigma}(1) = 0\) in
the reference implementation. Equation (13) implies the
post-perturbation target mean

\[\mu_{1,S}^{(d)} = c_{S} + D_{\mu}(d)\left( \mu_{1,S} - c_{S} \right),\quad\quad(14)\]

and target covariance

\[\Sigma_{1,SS}^{(d)} = D_{\sigma}(d)\Sigma_{1,SS}D_{\sigma}(d)^{\top}.\quad\quad(15)\]

Because non-target source symptoms are left unchanged, their covariance
remains \(\Sigma_{1,KK}\), while target--non-target cross-covariances
are attenuated only through the target scale map. Ordering the source
vector as \((K,S)\) gives

\[\Sigma_{1}^{(d)} = \begin{bmatrix}
\Sigma_{1,KK} & \Sigma_{1,KS}D_{\sigma}(d)^{\top} \\
D_{\sigma}(d)\Sigma_{1,SK} & D_{\sigma}(d)\Sigma_{1,SS}D_{\sigma}(d)^{\top}
\end{bmatrix}.\quad\quad(16)\]

Substituting the perturbed source moments into the longitudinal
transition gives

\[\mu_{2}^{(d)} = a + B\mu_{1}^{(d)},\quad\quad(17)\]

\[\Sigma_{2}^{(d)} = B\Sigma_{1}^{(d)}B^{\top} + \Psi.\quad\quad(18)\]

The mean downstream improvement induced by perturbing \(S\) is therefore

\[R_{S}(d) = \mu_{2} - \mu_{2}^{(d)} = B\left( \mu_{1} - \mu_{1}^{(d)} \right).\quad\quad(19)\]

Equation (19) is the central longitudinal propagation identity. It shows
explicitly that the effect is jointly determined by the amount of
source-state change and by the fitted transition operator. The
corresponding covariance equation (18) is retained because a
location-only intervention and a linked location--scale intervention can
have identical mean responses but different uncertainty and
bounded-scale behaviour.

\subsection{9.4 Temporal virtual
knockout}\label{temporal-virtual-knockout}

For one target \(i\), temporal virtual knockout (t-vKO) is the unit-dose
endpoint of the state operator. In the reference implementation,

\[E\left( X_{1,i}^{vKO} \right) = c_{i},\quad\quad Var\left( X_{1,i}^{vKO} \right) = 0.\quad\quad(20)\]

Let \(\delta_{i} = \mu_{1,i} - c_{i}\) be the modifiable distance from
the observed source mean to the anchor. Because only coordinate \(i\)
changes in the source mean, equation (19) reduces to

\[R_{i}^{vKO} = B_{\cdot i}\,\delta_{i},\quad\quad(21)\]

where \(B_{\cdot i}\) is column \(i\) of \(B\). The outcome-specific
standardised improvement is

\[\Delta_{j \leftarrow i}^{vKO} = \frac{B_{ji}\delta_{i}}{s_{2,j}},\quad\quad(22)\]

where \(s_{2,j} = \sqrt{\Sigma_{2,jj}}\) is the unperturbed follow-up
standard deviation. Equation (22) makes the distinction from outgoing
centrality explicit: a large \(\left| B_{ji} \right|\) does not
guarantee a large perturbation effect when \(\delta_{i}\) is small, and
a large source burden does not guarantee leverage when the outgoing
transition profile is weak or sign-cancelling.

An exact t-vKO does not require re-estimation of \(B\). The fixed source
value is passed through the already fitted transition equation. If a new
CLPN were instead refitted after exact knockout, the target predictor
would have zero variance and the refitted design matrix could become
rank deficient; that procedure answers a different question and was not
used here.

\subsection{9.5 Temporal virtual knockdown and dosage
perturbation}\label{temporal-virtual-knockdown-and-dosage-perturbation}

Temporal virtual knockdown (t-vKD) uses a partial dose \(d \in (0,1)\).
The linked reference map moves both the target location and residual
dispersion toward the anchor by the same proportion:

\[X_{1,i}^{(d)} = c_{i} + (1 - d)\left( X_{1,i} - c_{i} \right).\quad\quad(23)\]

Its mean and variance are

\[E\left( X_{1,i}^{(d)} \right) = c_{i} + (1 - d)\left( \mu_{1,i} - c_{i} \right),\quad\quad Var\left( X_{1,i}^{(d)} \right) = (1 - d)^{2}\sigma_{1,i}^{2}.\quad\quad(24)\]

The source mean improvement is \(d\delta_{i}\), hence

\[R_{i}(d) = dB_{\cdot i}\delta_{i}.\quad\quad(25)\]

If the downstream utility is a linear weighted functional of the mean
response,

\[G_{i}(d) = \frac{\sum_{j}^{}w_{j}\Delta_{j \leftarrow i}(d)}{\sum_{j}^{}w_{j}},\]

then equation (25) implies the exact linear null relation

\[G_{i}(d) = dG_{i}(1).\quad\quad(26)\]

Consequently,

\[\frac{G_{i}(d)}{d} = G_{i}(1),\quad\quad\left. \ \frac{\partial G_{i}(d)}{\partial d} \right|_{d = 0} = G_{i}(1).\quad\quad(27)\]

Thus efficacy, benefit per unit dose and local responsiveness are
mathematically redundant in an unbounded linear CLPN with a linear dose
map. A nonlinear dose-response curve is interpretable only when an
additional nonlinear element is specified.

For bounded symptom scales, one such source of nonlinearity is the
measurement boundary. Let \(h(y) = min\{ U,max(L,y)\}\) and
\(Y \sim N\left( m,s^{2} \right)\). With \(a = (L - m)/s\) and
\(b = (U - m)/s\), the bounded expectation is

\[E\{ h(Y)\} = L\Phi(a) + m\{\Phi(b) - \Phi(a)\} + s\{\phi(a) - \phi(b)\} + U\{ 1 - \Phi(b)\}.\quad\quad(28)\]

When \(s = 0\), the expression reduces to \(h(m)\). Equation (28)
provides an analytical way to distinguish genuine boundary-induced
curvature from coding artefacts in dosage analyses.

\subsection{9.6 Directed edge and source-node communication
blocking}\label{directed-edge-and-source-node-communication-blocking}

Communication blocking modifies the transition operator rather than the
source state. For a directed path \(i(T1) \rightarrow j(T2)\) and block
fraction \(q \in \lbrack 0,1\rbrack\), the edge-level operator is

\[B^{(i \rightarrow j,q)} = B - qB_{ji}e_{j}e_{i}^{\top}.\quad\quad(29)\]

For a prespecified source profile \(x^{ref}\), the predicted follow-up
difference between the unblocked and blocked models is

\[\Delta_{i \rightarrow j}^{edge}\left( q;x^{ref} \right) = \{ B - B^{(i \rightarrow j,q)}\} x^{ref} = qB_{ji}x_{i}^{ref}e_{j}.\quad\quad(30)\]

Averaging over a source distribution gives

\[E\left\{ \Delta_{i \rightarrow j}^{edge}\left( q;X_{1} \right) \right\} = qB_{ji}\mu_{1,i}e_{j}.\quad\quad(31)\]

Equation (31) explains why communication blocking must specify a
reference state. If predictors are mean-centred, \(\mu_{1,i} = 0\) and
the signed population-average prediction change is zero even when
\(B_{ji}\) is large. The present framework therefore uses either
clinically anchored profiles or individual-level predictions followed by
aggregation rather than interpreting coefficient magnitude itself as
intervention impact.

A full source-node block attenuates all outgoing transitions from source
\(i\):

\[B^{(i,q)} = B - qB_{\cdot i}e_{i}^{\top}.\quad\quad(32)\]

A cross-symptom-only version preserves autoregressive persistence and
blocks only off-diagonal outgoing paths:

\[B_{cross}^{(i,q)} = B - q\left( B_{\cdot i} - B_{ii}e_{i} \right)e_{i}^{\top}.\quad\quad(33)\]

To quantify topology-level propagation independently of a single
reference state, the reference implementation uses a finite-horizon
unsigned propagation functional

\[Q_{H}(B) = \mathbf{1}^{\top}\left\{ \sum_{h = 1}^{H}\left( \gamma|B| \right)^{h} \right\}\mathbf{1},\quad\quad 0 < \gamma < 1.\quad\quad(34)\]

and the node-centred communication-block score is

\[R_{i}^{comm} = \frac{Q_{H}(B) - Q_{H}\left( B^{(i,q)} \right)}{Q_{H}(B)}.\quad\quad(35)\]

Because \(Q_{H}\) uses \(|B|\), equation (35) measures route capacity
rather than signed clinical benefit. Signed reference-state effects from
equation (30) should therefore be inspected alongside the topology score
when inhibitory paths are scientifically meaningful.

\subsection{9.7 Combination perturbation and the additivity
constraint}\label{combination-perturbation-and-the-additivity-constraint}

For targets \(i\) and \(k\), let \(d_{i}\delta_{i}e_{i}\) and
\(d_{k}\delta_{k}e_{k}\) denote their source mean improvements. The
joint response is

\[R_{\{ i,k\}}\left( d_{i},d_{k} \right) = B\left( d_{i}\delta_{i}e_{i} + d_{k}\delta_{k}e_{k} \right) = R_{i}\left( d_{i} \right) + R_{k}\left( d_{k} \right).\quad\quad(36)\]

For any linear utility calculated on a common outcome set, equation (36)
yields

\[G_{\{ i,k\}}\left( d_{i},d_{k} \right) = G_{i}\left( d_{i} \right) + G_{k}\left( d_{k} \right),\quad\quad NA_{ik} = G_{\{ i,k\}} - G_{i} - G_{k} = 0.\quad\quad(37)\]

Therefore, a standard linear two-wave CLPN cannot create statistical
synergy from independent state perturbations. tSymPerturb instead
defines \emph{incremental combination value} on a common comparison set
that excludes targets \(i\) and \(k\):

\[I_{ik} = G_{ik}^{( - ik)} - max\left\{ G_{i}^{( - ik)},G_{k}^{( - ik)} \right\}.\quad\quad(38)\]

For a prespecified partner set \(T_{i}\), the reference target-level
combination score is

\[R_{i}^{comb} = \frac{1}{\left| T_{i} \right|}\sum_{k \in T_{i}}^{}\max\left( 0,I_{ik} \right).\quad\quad(39)\]

The signed \(I_{ik}\) values should also be retained so that
antagonistic or harmful combinations are not hidden. A non-zero additive
interaction contrast in equation (37) can arise only after introducing a
nonlinear outcome transform, interaction term, state-dependent
coefficient, constraint or another non-additive operation.

\subsection{9.8 Multi-wave propagation and intervention sequence
recursion}\label{multi-wave-propagation-and-intervention-sequence-recursion}

For \(T \geq 3\) occasions, the transition model generalises to

\[X_{t + 1} = a_{t} + B_{t}X_{t} + \varepsilon_{t}.\quad\quad(40)\]

If a one-time source intervention at occasion \(t\) creates mean
improvement \(\delta_{t} = \mu_{t} - \mu_{t}^{\ast}\), the one-step
response is \(B_{t}\delta_{t}\). After \(h\) observed transitions,

\[R_{t + h} = B_{t + h - 1}B_{t + h - 2}\cdots B_{t}\delta_{t}.\quad\quad(41)\]

Under a stationary transition matrix \(B_{t} = B\),

\[R_{t + h} = B^{h}\delta_{t}.\quad\quad(42)\]

Equations (41) and (42) distinguish immediate leverage from persistent
longitudinal leverage. They also clarify that iterating a two-wave
estimate beyond the observed interval requires a stationarity assumption
rather than additional information in the original data.

Repeated interventions can be represented by comparing an unperturbed
mean trajectory \(\mu_{t}^{0}\) with a perturbed pre-intervention
trajectory \(\mu_{t}^{-}\). Let
\(\Delta_{t}^{-} = \mu_{t}^{0} - \mu_{t}^{-}\) and let intervention
\(u_{t}\) create an additional improvement before the next transition,
so that \(\Delta_{t}^{+} = \Delta_{t}^{-} + u_{t}\). The next-wave
difference is then

\[\Delta_{t + 1}^{-} = B_{t}\left( \Delta_{t}^{-} + u_{t} \right).\quad\quad(43)\]

This recursion is the appropriate object for genuine temporal sequence
simulation. Order dependence can arise when interventions occur at
different observed occasions, when \(B_{t}\) changes over time, when
\(u_{t}\) depends on the current state, or when nonlinear constraints
are imposed. A two-wave CLPN contains only one observed transition and
therefore cannot identify this repeated sequence without additional
assumptions.

A finite-horizon discounted utility is

\[U_{i}^{(H)} = \sum_{h = 1}^{H}\eta^{h - 1}\frac{w^{\top}R_{t + h,i}}{\sum_{j}^{}w_{j}},\quad\quad 0 < \eta \leq 1.\quad\quad(44)\]

For the separate two-wave \emph{target-acquisition} problem, let
\(\pi = \left( \pi_{1},\ldots,\pi_{L} \right)\) and
\(S_{m} = \{\pi_{1},\ldots,\pi_{m}\}\). The decision objective is

\[J(\pi) = \sum_{m = 1}^{L}\gamma^{m - 1}\left\{ U\left( S_{m} \right) - U\left( S_{m - 1} \right) \right\} - \sum_{m = 1}^{L}\lambda_{\pi_{m}}.\quad\quad(45)\]

Equation (45) orders additions to an intervention set under explicit
cost and discount assumptions; it should not be described as evidence
for biological treatment order.

\subsection{9.9 Longitudinal utility outcomes and
tVPPS}\label{longitudinal-utility-outcomes-and-tvpps}

For target \(i\) at dose \(d\), the standardised outcome-specific
improvement is

\[\Delta_{j \leftarrow i}(d) = \frac{\mu_{2,j} - \mu_{2,j}^{(i,d)}}{s_{2,j}}.\quad\quad(46)\]

The total downstream efficacy includes the autoregressive destination:

\[G_{i}^{all}(d) = \frac{\sum_{j = 1}^{p}w_{j}\Delta_{j \leftarrow i}(d)}{\sum_{j = 1}^{p}w_{j}}.\quad\quad(47)\]

The cross-symptom spillover estimand excludes the target's own follow-up
value:

\[G_{i}^{cross}(d) = \frac{\sum_{j \neq i}^{}w_{j}\Delta_{j \leftarrow i}(d)}{\sum_{j \neq i}^{}w_{j}}.\quad\quad(48)\]

For threshold \(\tau\), breadth is

\[R_{i}^{breadth} = \frac{1}{p - 1}\sum_{j \neq i}^{}\mathbf{1}\left\{ \Delta_{j \leftarrow i}(1) \geq \tau \right\}.\quad\quad(49)\]

Let \(\mathcal{M}_{r}\) be the set of symptoms in module \(r\), and let
\(m(i)\) denote the target module. Cross-module reach is

\[R_{i}^{module} = \frac{1}{M - 1}\sum_{r \neq m(i)}^{}\mathbf{1}\left\{ \frac{1}{\left| \mathcal{M}_{r} \right|}\sum_{j \in \mathcal{M}_{r}}^{}\Delta_{j \leftarrow i}(1) \geq \tau_{m} \right\}.\quad\quad(50)\]

The positive spillover fraction separates distributed benefit from
autoregressive target persistence:

\[R_{i}^{spill} = \frac{\sum_{j \neq i}^{}w_{j}max\{\Delta_{j \leftarrow i}(1),0\}}{\sum_{j = 1}^{p}w_{j}max\{\Delta_{j \leftarrow i}(1),0\} + 10^{- 12}}.\quad\quad(51)\]

Together with communication-block value from equation (35) and
combination value from equation (39), these quantities define the seven
reference utility dimensions: downstream efficacy, cross-symptom
spillover, breadth, cross-module reach, communication-block value,
combination value and spillover fraction. Dose efficiency and low-dose
responsiveness are not included as separate dimensions in the linear
reference tVPPS because equations (26)--(27) show that they are
algebraically redundant with efficacy.

Each raw dimension \(R_{im}\) is direction-aligned and normalised only
within the prespecified candidate set:

\[{score}_{im} = 100\frac{R_{im} - \min_{\ell}R_{\ell m}}{\max_{\ell}R_{\ell m} - \min_{\ell}R_{\ell m}}.\quad\quad(52)\]

If a dimension is constant, it is assigned the neutral value 50. The
temporal virtual perturbation priority score is

\[{tVPPS}_{i} = \frac{\sum_{m = 1}^{7}\omega_{m}{score}_{im}}{\sum_{m = 1}^{7}\omega_{m}},\quad\quad\omega_{m} \geq 0.\quad\quad(53)\]

Equal weights were used only for internal methodological verification.
Because normalisation is candidate-set dependent, tVPPS is a relative
within-analysis rank and is not directly transportable across cohorts,
networks or alternative candidate sets. Robustness and sampling
uncertainty are reported separately rather than treated as intervention
utility.

\subsection{9.10 Simulation generating
mechanism}\label{simulation-generating-mechanism}

The generating system contained 22 named synthetic symptoms organised
into four modules. Baseline \(X_{1}\) followed a multivariate normal
distribution with heterogeneous means and standard deviations. Its
correlation matrix was generated from one weak general factor and four
module factors so that within-module dependence was stronger than
between-module dependence. The transition matrix contained 22
autoregressive coefficients and 53 non-zero off-diagonal paths; 47
cross-lagged paths were positive and 6 were negative. Strong
prespecified bridge transitions included Insomnia\(\rightarrow\)Fatigue,
Fatigue\(\rightarrow\)Concentration difficulty,
Rumination\(\rightarrow\)Anxiety, Rumination\(\rightarrow\)Insomnia,
Anxiety\(\rightarrow\)Palpitations, Pain\(\rightarrow\)Insomnia and
Kinesiophobia\(\rightarrow\)Activity avoidance. Autoregressive
coefficients ranged from 0.43 to 0.71 and the spectral radius of \(B\)
was 0.827.

Residual errors were independent Gaussian variables with standard
deviations from 0.38 to 0.52. The intercept was selected so that the
unperturbed follow-up mean equalled the baseline mean. From equation
(9), this requires

\[a = (I - B)\mu_{1}.\quad\quad(54)\]

Baseline means ranged from 1.44 to 2.64 and baseline standard deviations
from 0.63 to 0.83. Population perturbation outcomes were calculated from
the known \(\mu_{1}\), \(\Sigma_{1}\), \(B\) and \(\Psi\) before any
finite samples were generated.

At each sample size \(n \in \{ 250,500,1000\}\), 200 independently
generated datasets were created from distinct child seeds of master seed
20260816. Within each dataset, T1 and T2 variables were standardised and
each T2 symptom was regressed on all T1 symptoms using lasso with
penalty \(\alpha = 0.03\) and no intercept on the standardised scale.
The sample-specific zero anchor was transformed using equation (12).
Network estimation, perturbation, utility calculation, candidate-set
normalisation and ranking were rerun from the beginning for every
independent dataset.

\subsection{9.11 Analytical, Monte Carlo and algebraic
verification}\label{analytical-monte-carlo-and-algebraic-verification}

For each of the 22 targets, 250,000 independent post-t-vKO draws were
generated from the known source distribution and residual distribution.
The Monte Carlo estimate of the outcome response was

\[{\widehat{R}}_{i,j}^{MC} = {\bar{X}}_{2,j}^{base} - {\bar{X}}_{2,j}^{vKO(i)}.\quad\quad(55)\]

It was compared with the analytical response in equation (21) after
standardisation by \(s_{2,j}\). These Monte Carlo draws were used only
to verify the analytical expectation and were not treated as independent
statistical units.

Two exact algebraic checks were applied to every implementation. First,
the unbounded linear reference model must satisfy equation (26) for
every target and dose. Second, every two-target state perturbation must
satisfy equations (36)--(37). Numerically meaningful deviations from
either identity indicate an implementation error or the introduction of
a nonlinear operation and should be diagnosed before interpreting target
rankings.

The stationary multi-wave stress test reused the known \(B\) for four
transitions and compared one-step efficacy with \(B^{4}\delta_{i}\) and
with the discounted cumulative utility in equation (44). The
repeated-intervention analysis applied equation (43) to the eight
highest population tVPPS candidates. Because the same transition matrix
was reused beyond the observed two-wave interval, these results were
interpreted as strategy simulations under stationarity rather than
identified treatment sequences.

\subsection{9.12 Finite-sample recovery, benchmark comparisons and
uncertainty}\label{finite-sample-recovery-benchmark-comparisons-and-uncertainty}

The independent statistical unit for finite-sample verification was an
independently generated dataset. The primary recovery statistic was
Spearman correlation between estimated and population tVPPS ranks. We
additionally calculated top-five recovery and RMSE on the 0--100 tVPPS
scale. Medians and 2.5th--97.5th percentile Monte Carlo intervals
summarised the 200 replicate distributions at each sample size.

To determine whether tVPPS simply reproduced a standard longitudinal
centrality measure, the population ranking was compared with absolute
outgoing strength

\[S_{i}^{out} = \sum_{j = 1}^{p}\left| B_{ji} \right|,\quad\quad(56)\]

signed outgoing expected influence

\[EI_{i}^{out} = \sum_{j = 1}^{p}B_{ji},\quad\quad(57)\]

and a baseline-burden-only rule based on \(\mu_{1,i}\). Rank correlation
and top-five overlap were used as descriptive benchmark losses.

For applied data, uncertainty should be propagated through the complete
pipeline. In bootstrap replicate \(b\), let \({\widehat{r}}_{i}^{(b)}\)
be the resulting rank of target \(i\). Rank uncertainty and top-\(K\)
selection probability can be summarised as

\[P_{i}^{(K)} = \frac{1}{B_{boot}}\sum_{b = 1}^{B_{boot}}\mathbf{1}\left\{ {\widehat{r}}_{i}^{(b)} \leq K \right\}.\quad\quad(58)\]

The bootstrap should repeat participant resampling, standardisation,
anchor transformation, CLPN estimation, perturbation, utility
calculation, normalisation and ranking. Robustness is therefore treated
as an uncertainty diagnostic rather than as an eighth utility component.

\subsection{9.13 Required validation
extensions}\label{required-validation-extensions}

The current numerical experiment is an internal verification under a
model-compatible continuous Gaussian CLPN. A general methodological
claim requires factorial stress tests that vary data type (continuous,
ordinal and zero-inflated), floor and ceiling effects, measurement
error, latent common causes, missingness and attrition, network density,
edge-sign balance, \(p/n\) ratio, estimator and penalty selection,
time-varying coefficients, nonlinear transitions and participant-level
heterogeneity. These settings should distinguish failure caused by
network estimation from failure caused by the perturbation operator
itself.

\subsection{9.14 Software and
reproducibility}\label{software-and-reproducibility}

The validation was implemented in Python using NumPy, pandas, SciPy and
scikit-learn; figures were generated with Matplotlib. All data in the
methodological verification are synthetic and use master seed 20260816.
Before submission, the generating parameters, validated source code,
machine-readable target scores and figure source data should be archived
in a public version-controlled repository with a persistent release
identifier. An independent implementation in the intended analysis
environment should reproduce all analytical identities and source-data
tables.

\section*{Data availability}\label{data-availability}

All data used in the present methodological verification are synthetic.
The generating parameters, example datasets and source data underlying
the validation figures should accompany the public reproducibility
bundle. No patient data were used in the simulation results reported
here.

\section*{Code availability}\label{code-availability}

The core computational code underlying the original SymPerturb
implementation has been published previously in a related methodological
article. The tSymPerturb implementation extends that framework to
directed transition matrices, temporal state operators, directed
communication blocking, longitudinal utility outcomes and multi-wave
strategy simulation. The validated longitudinal code and
machine-readable outputs should be released in a version-controlled
public repository upon publication.

\section*{Acknowledgements}\label{acknowledgements}

This work was supported by the National Natural Science Foundation of
China (Grant No. 72574043), the Shanghai Pujiang Program (Grant No.
24PJC014), and the China University Industry--Research Innovation
Fund--Digital Intelligence Innovation and Talent Program (Grant No.
2024LC007), all awarded to Z.Z. The funders had no role in study design,
data generation, analysis, interpretation, decision to publish or
preparation of the manuscript.

\section*{Author contributions}\label{author-contributions}

Z.Z. conceived and developed the tSymPerturb framework, designed the
methodological study and drafted the manuscript. J.Y., T.H., Z.Y. and
J.W. provided methodological and conceptual feedback and critically
reviewed the framework. All authors will review and approve the final
version.

\section*{Competing interests}\label{competing-interests}

The authors declare no competing interests.

\end{document}